\documentclass[prl,showpacs,superscriptaddress,preprintnumbers,amssymb,twocolumn,amsfonts]{revtex4}
\usepackage{graphicx}
\usepackage{amssymb}
\usepackage{amsmath}
\usepackage{subfigure}
\usepackage{bm}

\usepackage[usenames,dvipsnames]{color}         

\begin{document}

\input epsf
\def\beq{\begin{equation}}
\def\eeq{\end{equation}}
\def\beqn{\begin{eqnarray}}
\def\eeqn{\end{eqnarray}}
\newcommand{\etal}{{\textit{et al.}}}
\def\ket#1{\vert #1 \rangle}
\def\bra#1{\langle #1 \vert}
\def\ev#1{\langle #1 \rangle}
\def\ip#1#2{\langle #1 \vert #2 \rangle}
\def\me#1#2#3{\langle #1 \vert #2 \vert #3 \rangle}
\renewcommand{\bf}{\mathbf}

\title{Bound states and $\textrm{E}_8$ symmetry effects in perturbed quantum Ising chains}

\author{Jonas A.~Kj\"{a}ll}
\affiliation{Department of Physics, University of California,
Berkeley, CA 94720}
\author{Frank Pollmann}
\affiliation{Department of Physics, University of California,
Berkeley, CA 94720}
\author{Joel~E.~Moore}
\affiliation{Department of Physics, University of California,
Berkeley, CA 94720} \affiliation{Materials Sciences Division,
Lawrence Berkeley National Laboratory, Berkeley, CA 94720}
\date{January 15, 2011}

\begin{abstract}
In a recent experiment on $\mathrm{CoNb}_2\mathrm{O}_6$, Coldea \etal~\cite{Coldea} found for the first time experimental evidence of the exceptional Lie algebra 
$\textrm{E}_8$.  The emergence of this symmetry was theoretically predicted long ago for the transverse quantum Ising chain in the presence of a weak longitudinal field.  We consider an accurate microscopic model of $\mathrm{CoNb}_2\mathrm{O}_6$ incorporating additional couplings and calculate numerically the dynamical structure function using a recently developed matrix-product-state method.  The excitation spectra show bound states characteristic of the weakly broken $\textrm{E}_8$ symmetry.  We compare the observed bound state signatures in this model to those found in the transverse Ising chain in a longitudinal field and to experimental data.
\end{abstract}
\maketitle

The one-dimensional (1D) quantum Ising model in transverse and longitudinal fields is one of the most studied theoretical models in condensed matter physics.  It is a relatively simple model that contains very rich physics; for example, it contains a quantum critical point (QCP) at zero longitudinal field related to the 2D classical Ising model.  A remarkable fact is that the integrability present at the critical point remains under addition of a longitudinal field as a mass-generating perturbation.  Zamolodchikov conjectured in 1989 an S-matrix describing eight emergent particles whose mass ratios are connected to the roots of the Lie algebra $E_8$~\cite{Zamolodchikov,Mussardo}.  Recently, Coldea et al.  performed neutron scattering experiments on $\textrm{CoNb}_2\textrm{O}_6$ (cobalt niobate), a material that to a good approximation can be described by a quantum Ising chain.  At low temperatures and in the presence of a strong external transverse magnetic field which tunes the system to near criticality, the observed spectrum shows characteristic excitations of the  $E_8$ symmetry~\cite{Coldea}. 

However, a serious problem in comparing theory and experiment is that the real
material has additional couplings that strictly speaking invalidate the exact
solution, and until recently it was impractical to extend the theory
non-perturbatively to include these couplings.  In this Letter, we study a theoretical model for $\mathrm{CoNb}_2\mathrm{O}_6$ which includes in addition to the Ising interaction other interactions arising from the lattice structure and the weak coupling between the chains.   Using this model, we calculate the dynamical spectral function and compare the results to the observed spectra.  Close to the QCP,  the model retains features expected from the quantum Ising model, in particular the characteristic particles of the $\textrm{E}_8$ symmetry. 

\begin{figure}[htbp]
  \begin{center}
    \subfigure{\includegraphics[width=42mm]{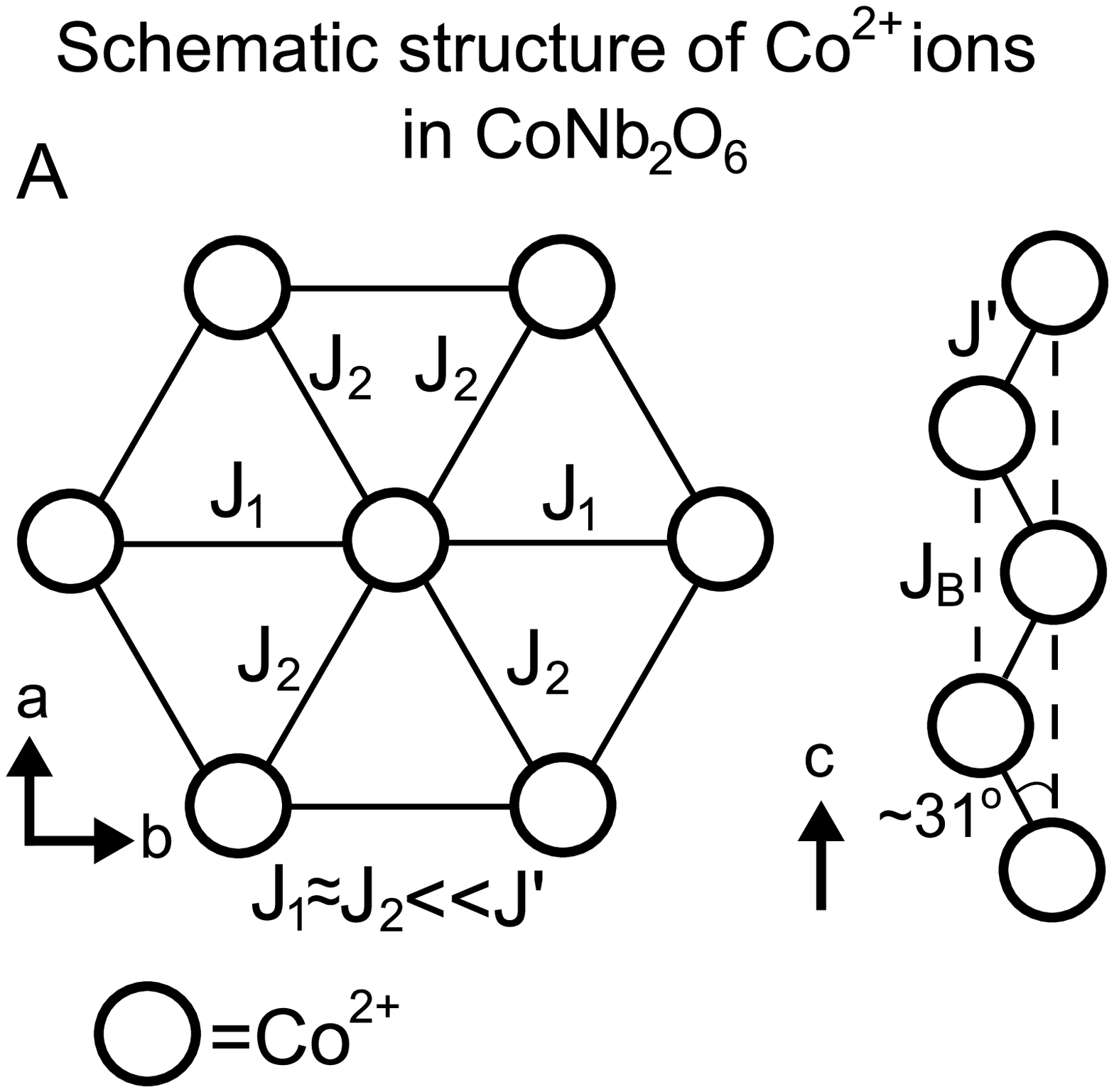}}
      \subfigure{\includegraphics[width=38mm]{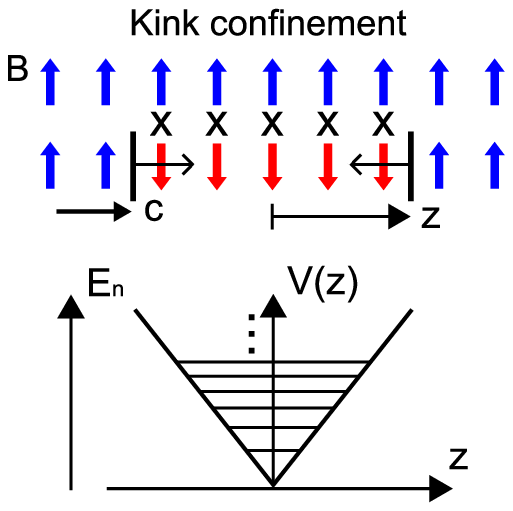}}\\
        \subfigure{\includegraphics[width=42mm]{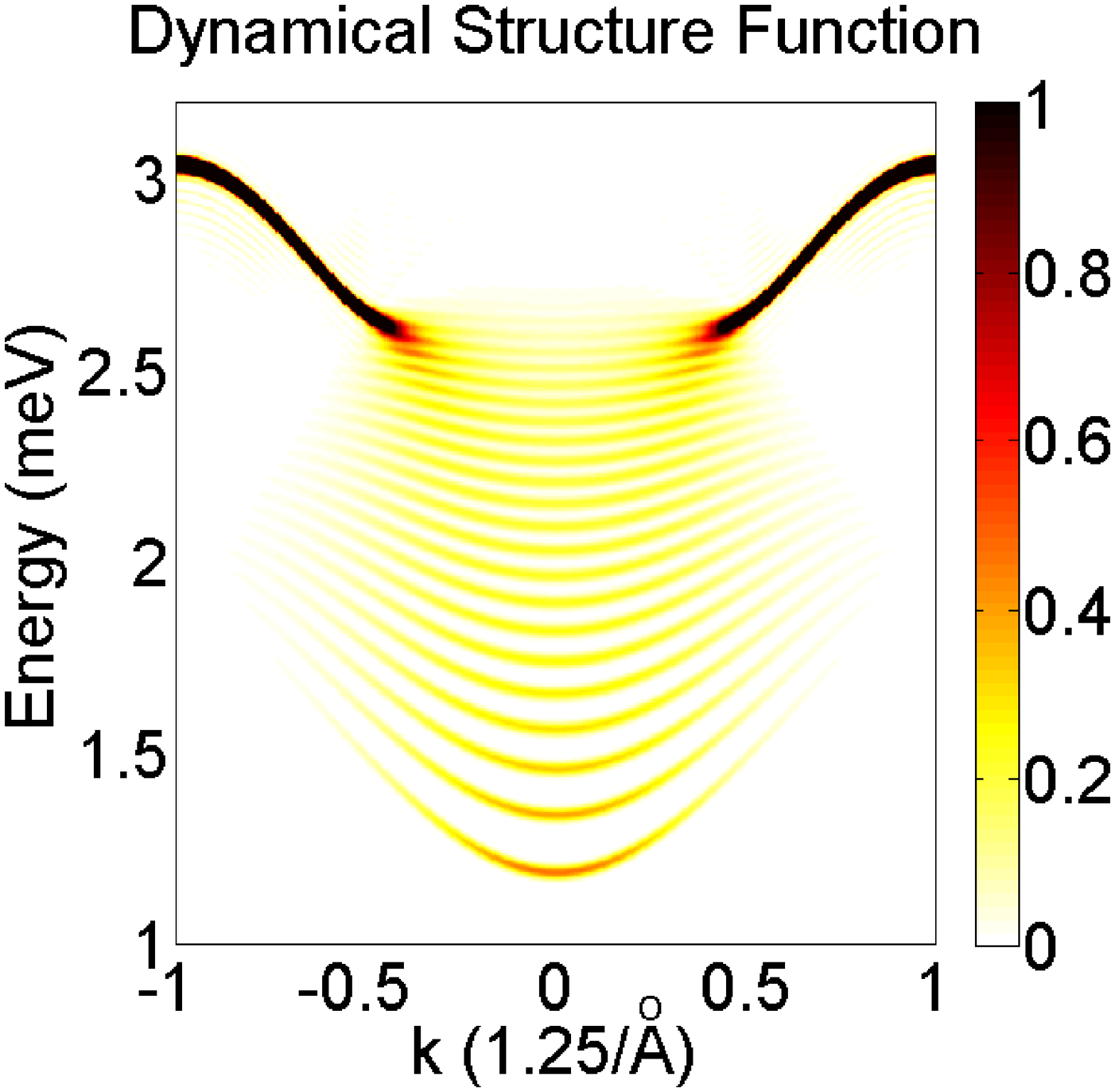}}
          \subfigure{\includegraphics[width=42mm]{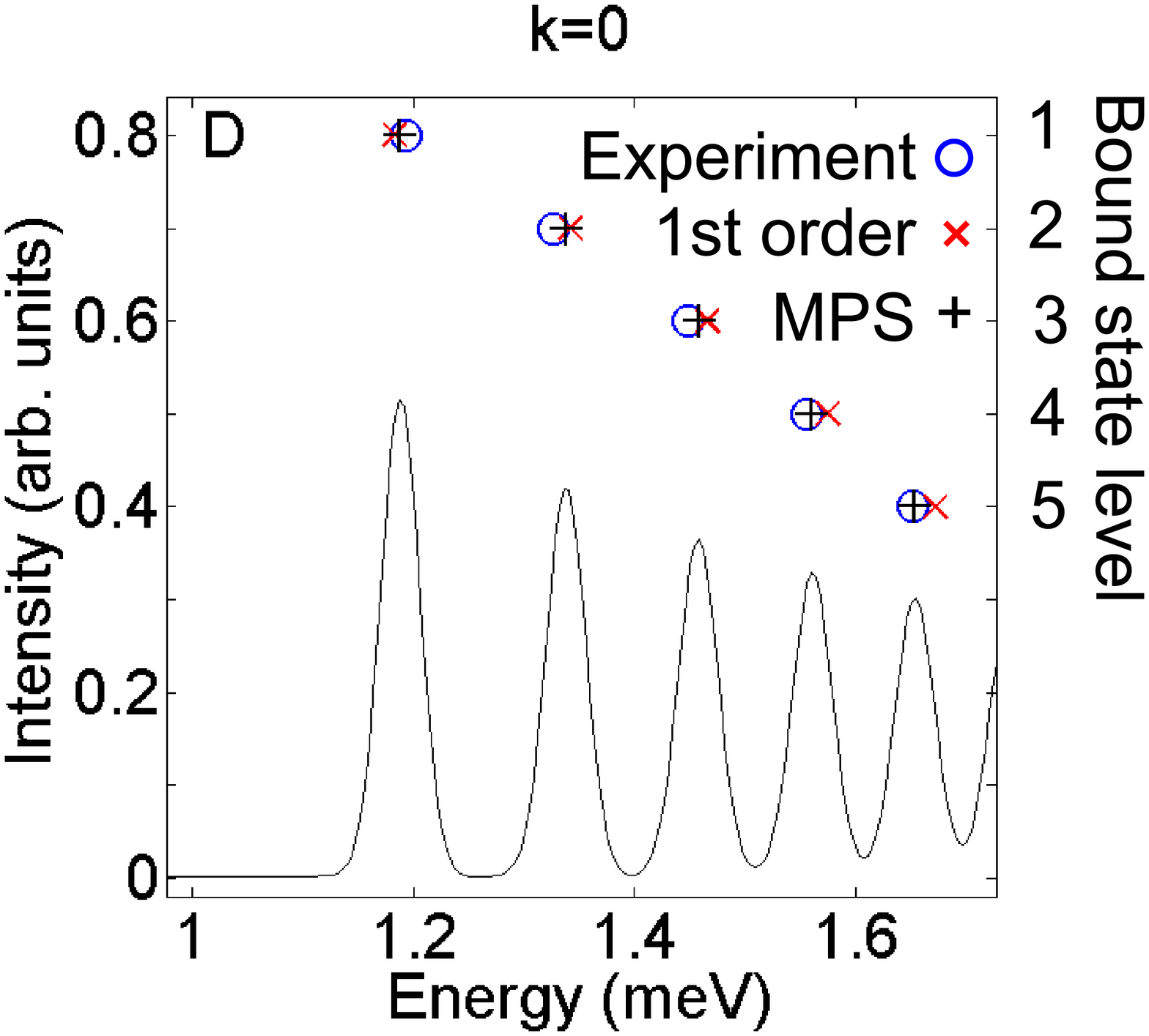}}

    \caption{(Color online) (A) Ising spins on the $Co^{2+}$ ions are strongly coupled in 1D along zig-zag chains. The $Co^{2+}$ ions are ordered in a weakly coupled triangular lattice in the plane perpendicular to the chain direction, with \textbf{a}, \textbf{b} and \textbf{c} orthogonal unit vectors.  (B) Confinement of kinks: the potential energy between kinks increase linearly, along the $z$ coordinate in the \textbf{c} direction, as more interchain bonds turn energetically unfavorable.  The energy levels are given by the negative zeros of the Airy function. (C-D) The full Hamiltonian describing $\textrm{CoNb}_2\textrm{O}_6$ at no external magnetic field.  (C) The dynamical structure function.  (D) The cross section of (C) at zero momentum showing the masses of the first five bound states.  A comparison of these masses from our MPS calculations (pluses) with the experimental results (circles) and the exact solution of the proposed first order phenomenological model (crosses).}
    \label{1}
  \end{center}
\end{figure}

We begin by  deriving the theoretical model used to describe the low-energy physics of $\mathrm{CoNb}_2\mathrm{O}_6$.  The spin lattice structure consists of chains of easy axis spins, realizing a two level system, on the $\mathrm{Co}^{2+}$ ions coupled by a ferromagnetic Ising interaction along the chain direction, see Fig.~\ref{1}A.  We thus start from the quantum Ising chain, described by the Hamiltonian
\begin{equation}\label{QIC}
H=-J\displaystyle\sum_n S_n^zS_{n+1}^z-h^x\displaystyle\sum_n S_n^x
\end{equation}
where $J>0$ favors a ferromagnetic state ($|\uparrow\uparrow ...\uparrow\rangle$ or $|\downarrow\downarrow ...\downarrow\rangle$).  When the transverse field is increased past the QCP $|h_c^x|=J/2$, the system undergoes a phase transition into a paramagnetic state $|\rightarrow\rightarrow ...\rightarrow\rangle$.  This model is exactly solvable using a Jordan-Wigner transformation which transforms the spins into non-interacting fermions~\cite{Sachdev}.

The lowest lying excitation energy is similar on both sides of the QCP due to the self-duality of the model and goes to zero, that is, the gap closes, at the QCP.   However, the double degeneracy of the ferromagnetic ground state leads to a fractionalization of the experimental excitation, a spin flip, into two freely moving domain walls or kinks. 
We now take into account terms which result from the three-dimensional (3D) lattice structure of $\mathrm{CoNb}_2\mathrm{O}_6$.  A recent theoretical study by Lee \etal\ investigates a three-dimensional model of $\textrm{CoNb}_2\textrm{O}_6$~\cite{LeeBalents}. They show that the plane perpendicular to the chain, a weakly coupled triangular lattice, see Fig.~\ref{1}A, has ferrimagnetic order to transverse field strengths well passed $h_c^x$.  The interchain couplings in a 3D magnetic ordered material at low temperature can be well approximated by a chain in a local effective longitudinal field $h^z=\sum_{\delta}J_{\delta}\langle S^z\rangle$ with the sum over all nearest interchain bonds~\cite{Carr}.  This field favors the ferromagnetic phase, breaks its two-fold symmetry and moves the system away from the QCP.  It also splits up the continuum into bound states by confining the kinks. 
At low transverse field and small bound state momentum, this can be described by a one-dimensional Schr\"{o}dinger equation with a linear confining potential with the energy levels given by the negative zeros of the Airy function, see Fig.~\ref{1}B~\cite{McCoy}.  This solution has later been extended to all possible bound state momenta~\cite{Rutkevich}.  Close to the QCP $(h^x=h^x_c,|h^z|\ll |h^x_c|)$, the eight massive particles described by the $\textrm{E}_8$ symmetry can be seen either as asymptotic states or as bound states of a pair of particles of this theory~\cite{Zamolodchikov,Mussardo}.

Although $\textrm{CoNb}_2\textrm{O}_6$ to a good approximation can be described by a quantum Ising chain, a realistic model must contain more interactions~\cite{Coldeasup}.  It has a strong easy axis character, but a weak XX part is still present.  The chains have a zig-zag structure, making the next-nearest neighbor (nnn) interaction important as well, see Fig.~\ref{1}A.  The measured Ising exchange energy $J$ is unusually low, likely due to a competition from an antiferromagnetic nnn interaction.  Taking into account all terms, the resulting Hamiltonian reads
\begin{eqnarray}\label{ph}
H&=&-J'\displaystyle\sum_n S_n^zS_{n+1}^z-h^x\displaystyle\sum_n S_n^x-h^z\displaystyle\sum_n S_n^z\\
&-&J_p\displaystyle \sum_n\left(S_n^xS_{n+1}^x+S_n^yS_{n+1}^y\right)+J_B\displaystyle\sum_n S_n^zS_{n+2}^z. \nonumber
\end{eqnarray} 
The numerical values for the coupling constants to describe $\textrm{CoNb}_2\textrm{O}_6$ are obtained by matching the experimental neutron scattering intensity at zero applied transverse field with our numerical calculations.  We compare the dynamical structure function $S^y(k,\omega)$, the Fourier transform of the dynamic two-point correlations
\begin{equation}\label{coor}
C^y(n,t)=\langle \psi_0|S_n^y(t)S_0^y(0)|\psi_0 \rangle.
\end{equation}  

For the numerical calculations, we use the \emph{time evolving block decimation} (TEBD)~\cite{Vidal-2003a,Vidal-2007} method which provides an efficient method to perform a time evolution of quantum states in one-dimensional systems.  The evolution of a random state of an infinite chain in imaginary time is used to calculate the ground state $|\psi_0\rangle$ and an evolution in real time allows us to calculate the dynamic two-point correlations directly.  The TEBD algorithm can be seen as a descendant of the \emph{density matrix renormalization group}~\cite{white2} method and is based on a matrix product state (MPS) representation~\cite{Fannes-1992,ostlundrommer} of the wavefunctions.  Algorithms of this type are efficient because they exploit the fact that the ground-state wave functions are only slightly entangled, especially away from criticality~\cite{GottesmanHastings}. As the entanglement grows linearly as a function of time, the simulations of long time evolutions is numerically very difficult.  To be able to simulate long enough times and thus to get sufficiently good energy resolution in the calculated spectral functions, we use a number of methods to accelerate the time evolution. We use linear predictions to extrapolate the dynamical correlation functions to very long times~\cite{Barthel-2009,White-2008} and take advantage of the ``light-cone'' like spread of the entanglement by adding more sites to the chain as time increases. As the calculation of the correlation functions $C^y(n,t)$ is numerically very expensive,  we calculate it only for certain time steps and then interpolate its values. In order to estimate the errors of our simulations, we calculate the truncation error, i.e., the truncated weight of the wave function at a time step, which gives an upper bound for the  truncation effects on local expectation values ($ \lesssim 10^{-6}$ for all simulations presented in this paper).  In addition we also checked the dependence of the measured observables on the matrix dimension $\chi$ per site and the time steps $\Delta t$, settling for $\chi=45$ and $\Delta t=0.04$ meV$^{-1}$ for the simulations presented in this paper.

 The numbers we use are $J'=J+J_B=2.43$ meV, $h^x=0.354$ meV, $h^z=0.035$ meV, $J_p=0.52$ meV and $J_B=0.60$ meV, see Fig.~\ref{1}C and compare it to Fig. 3 in Ref.~\onlinecite{Coldea}.  A cross-section with the bound state ``masses'' (i.e., the energies of the bound states at zero momentum) is presented in Fig.~\ref{1}D with the experimentally measured masses and Rutkevich's exact solution of Coldea \etal~first order model for reference in Fig.~\ref{1}D~\cite{Coldea,Rutkevich2}.  Note that our full Hamiltonian agrees to first order in perturbation theory with the phenomenological model used there, although the coupling constant for that model is slightly larger than ours~\cite{Coldeasup}.
\begin{figure}[htbp]
  \begin{center}
    \subfigure{\includegraphics[width=42mm]{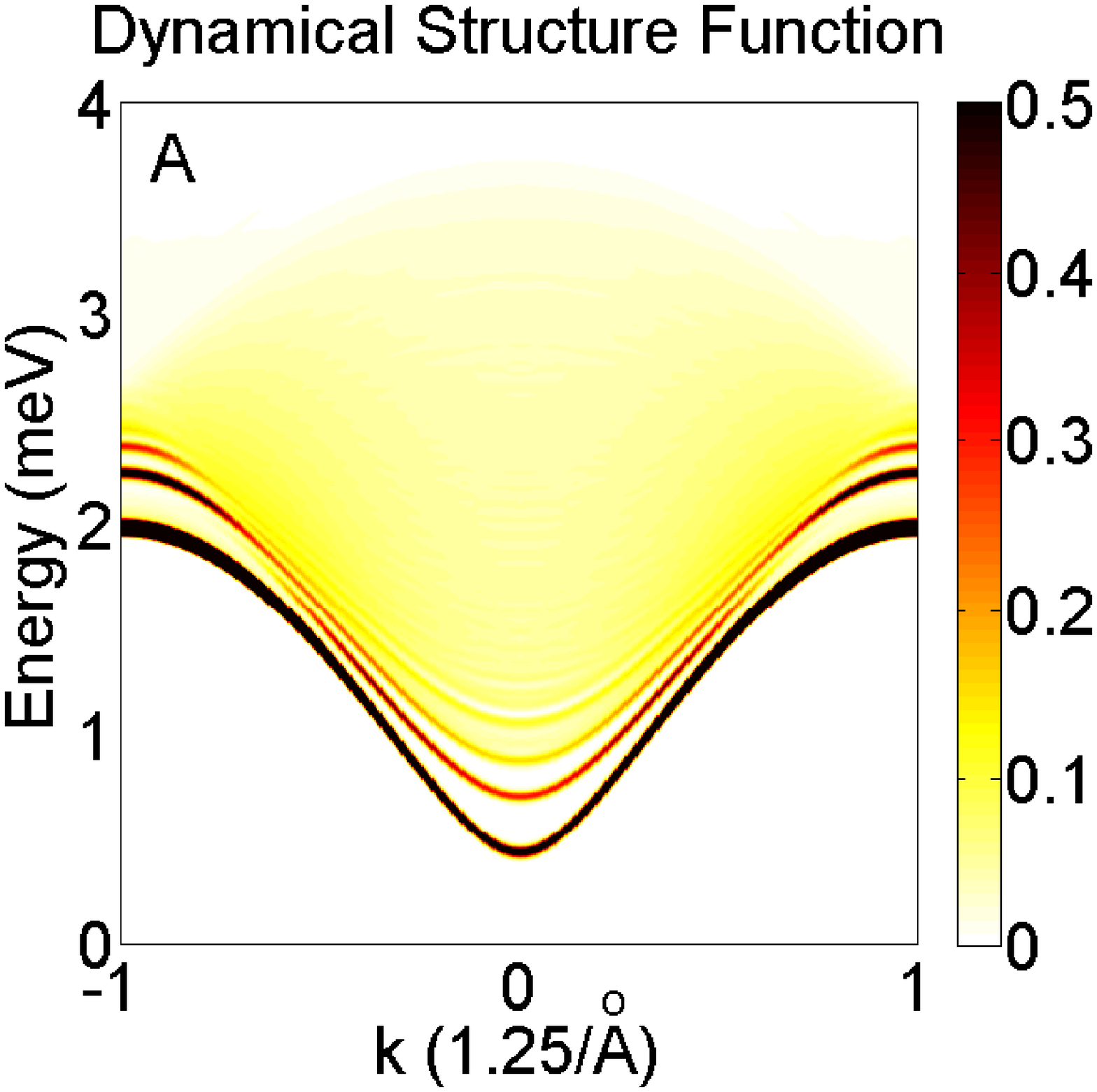}}
        \subfigure{\includegraphics[width=42mm]{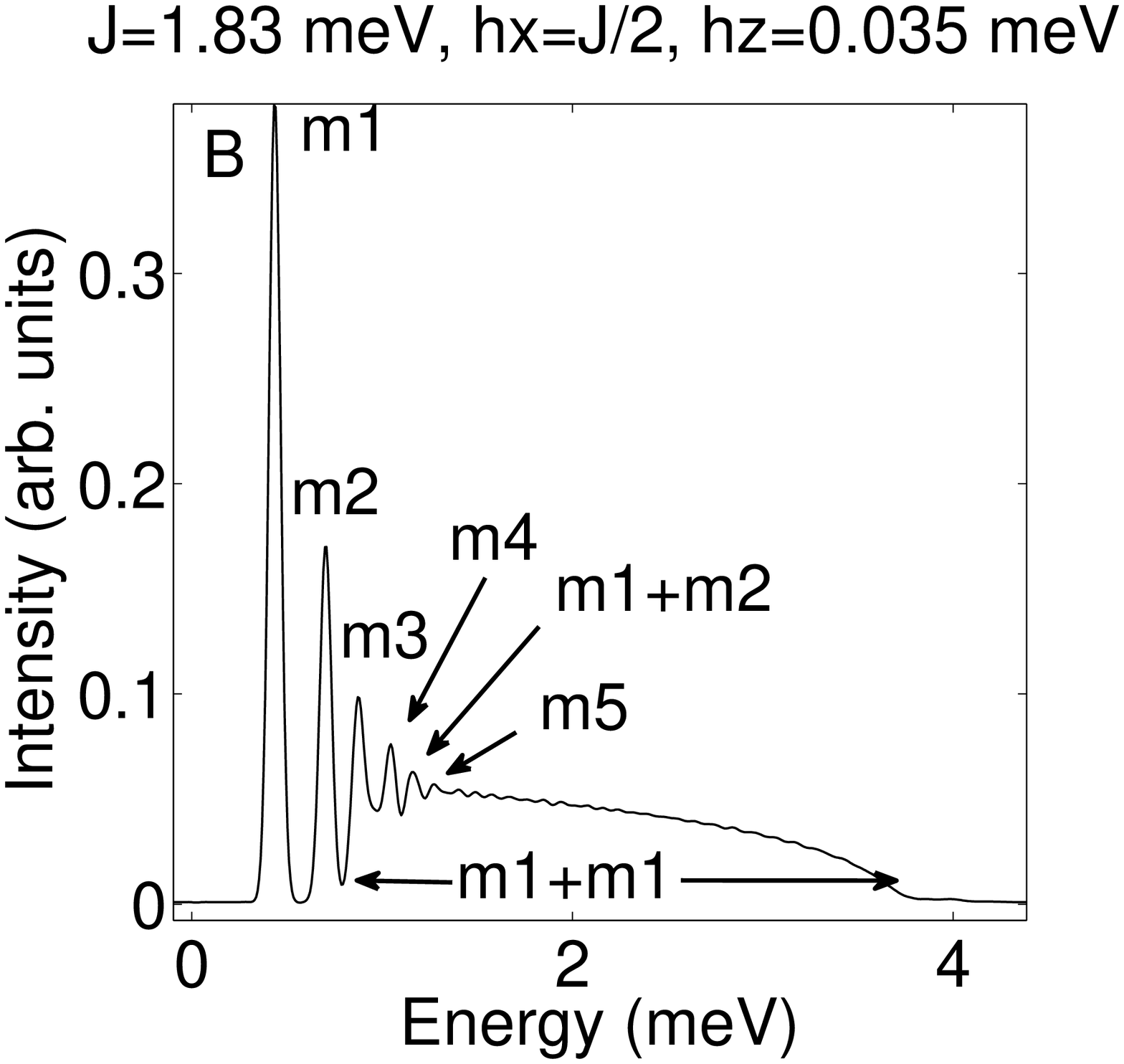}}\\
            \subfigure{\includegraphics[width=42mm]{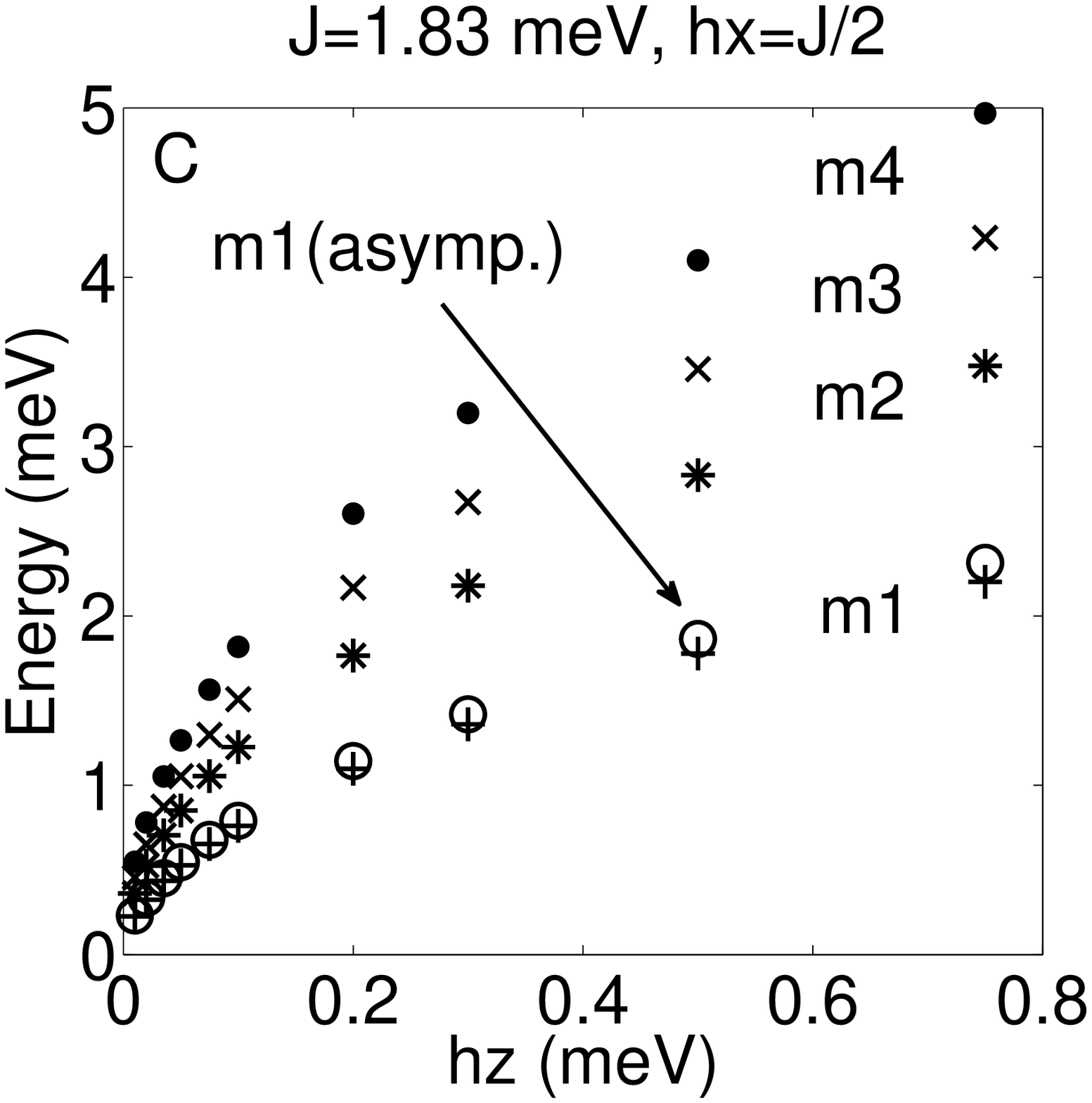}}
                \subfigure{\includegraphics[width=42mm]{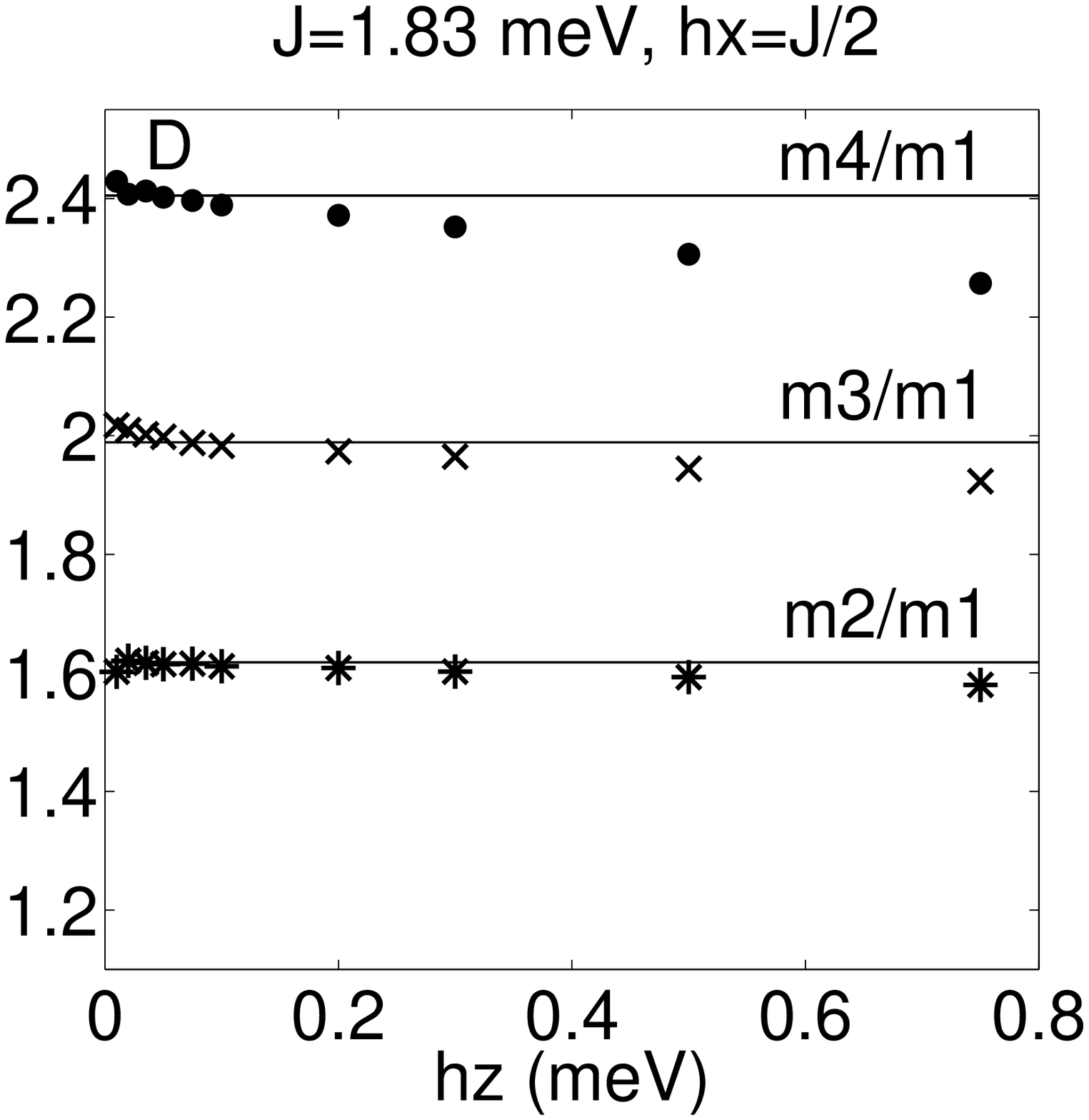}}\\
                    \subfigure{\includegraphics[width=42mm]{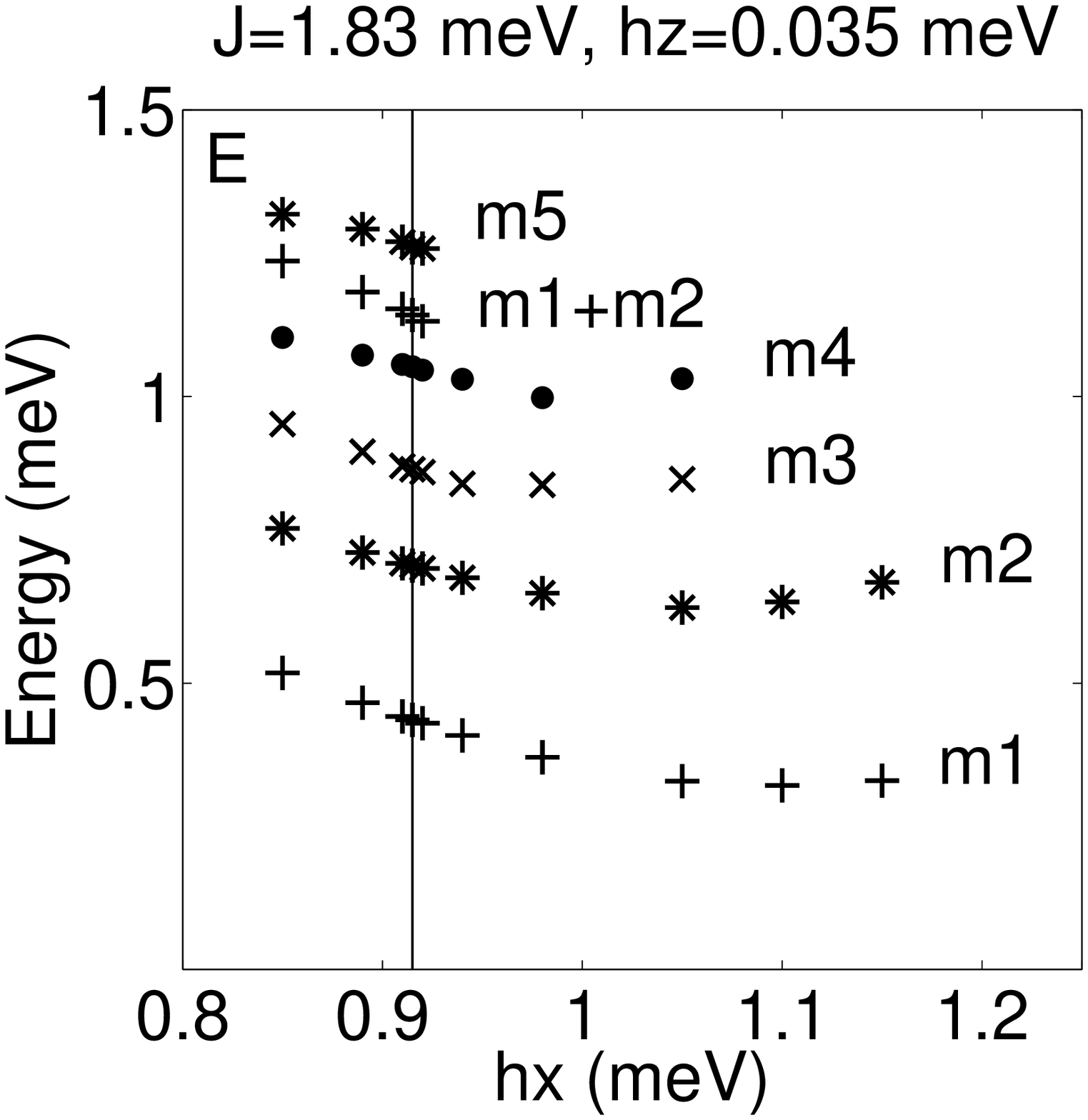}}
        \subfigure{\includegraphics[width=42mm]{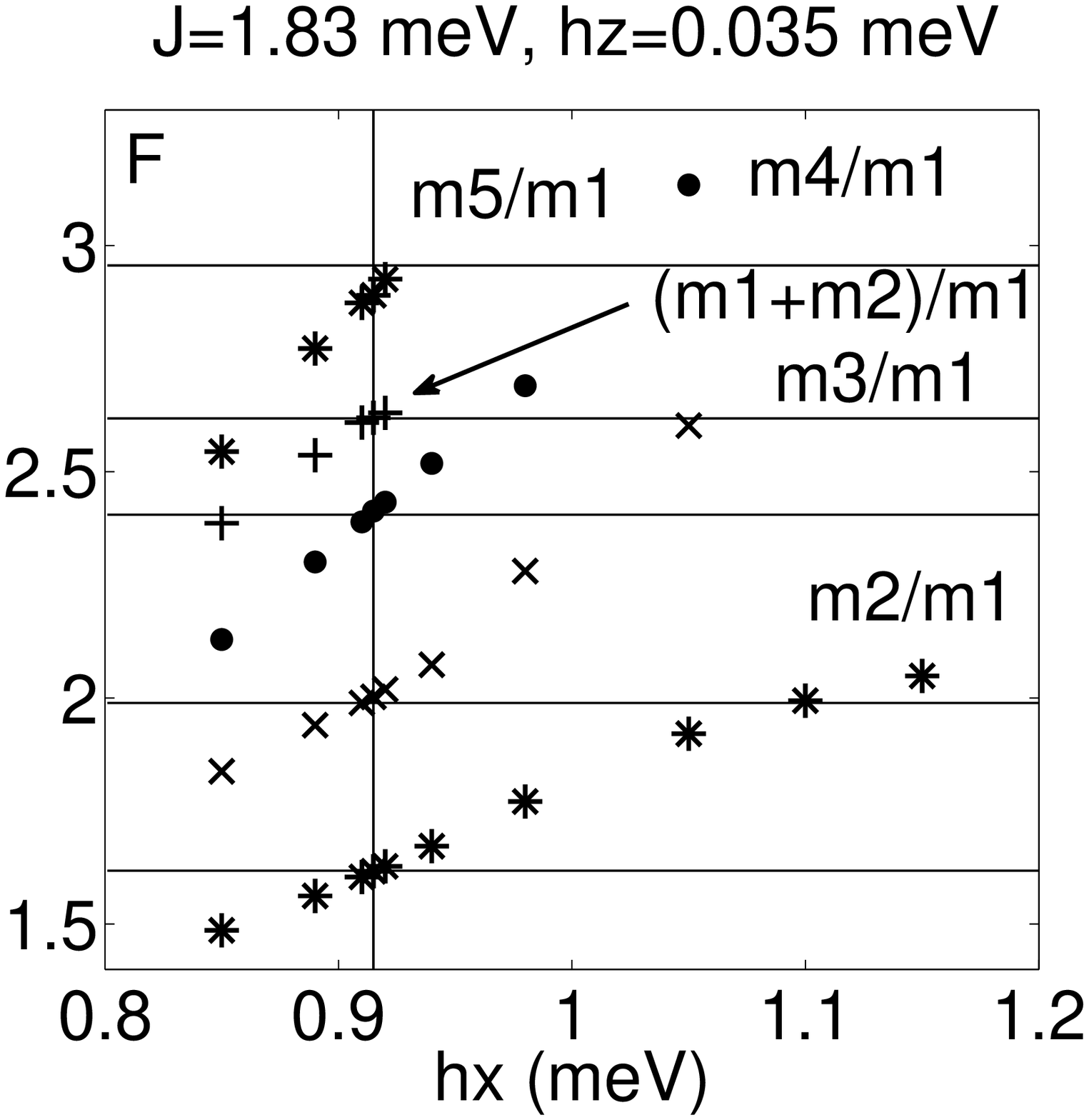}}
    \caption{(Color online) Ising chain in a transverse and a longitudinal field.  (A) The dynamical structure function at $h^x=h^x_c$ and $h^z=0.035$ meV.  (B) The cross section of (A) at zero momentum.  The five lowest bound states and two bound state pairs can be distunguished.  (C) ($h^x=h^x_c$) The mass of the four lowest bound states and an asymptotic expansion for $m_1$ from $h^z\rightarrow 0$ as a function of the longitudinal field.  (D) ($h^x=h^x_c$) Relative mass of the lowest bound states compared to the analytical predicted values at $h^z\rightarrow 0$.  (E) ($h^z=0.035$ meV) The bound state masses as a function of the transverse field ($h^x$).  The energy gap is smallest around $h^x=1.10$ meV, however the minimum in the higher bound states occurs for lower fields.  (F) ($h^z=0.035$ meV) The relative masses for the bound states increases roughly linear as a function of the field ($h^x$) and passes the analytical calculated values at $h_c^x$. }
    \label{2}
  \end{center}
\end{figure} 

Before investigating this model near the QCP, we start with a pure quantum Ising chain where the $\textrm{E}_8$ symmetry is expected to be present ~\cite{Zamolodchikov}.  The dynamical structure function is calculated over the whole Brillouin zone (for various parameters $J=1.83$ meV, $h^x$, $h^z$), see Fig.~\ref{2}A for an example and we focus on the cross section at zero momentum where some comparison with earlier work can be done.  Fig.~\ref{2}B is an example of these cross sections at $h^x=h^x_c$ with $h^z=0.035$ meV.  The lowest four bound states can be easily detected and one or two more can be distinguished.  Bound state pairs $m_1+m_1$ (overlapping with $m_3$) and $m_1+m_2$ have similar intensity to the nearby bound states, making both types simultaneously observable.  They are created in a ``spinon jet'', where the two kinks (also known as spinons) in a bound state have been stretched far enough apart to make it energetically favorable to flip a spin between them to form two more kinks that each form an independent low energy bound state with one of the original kinks.  The independent motion of these two bound pairs appears as a continuum in the dynamical structure function.  This process is reminiscent of quark dynamics, where the quarks are confined and cannot be isolated singularly. Finding condensed matter analogues of confinement effects known from high energy physics might help us to improve our understanding of underlying mechanisms; see e.g. Lake \etal~\cite{Lake}.

The weight of the continuum decreases with increasing longitudinal field; this is also the case for the weight of the higher bound states but to a lesser extent.  The gap and the spacing between the bound states increase with increasing longitudinal field; see Fig.~\ref{2}C where data from more simulations are presented, together with an asymptotic expansion from the exact analytical limit ($h^x=h^x_c$, $h^z\rightarrow 0$) of the lowest bound state.  The analytical expression for the gap is $m_1\approx CJ/4 (2h^z/J)^{8/15}$, with $C=4.40490858/0.7833$, showing good agreement with our results to high longitudinal field strengths~\cite{Zamolodchikov,Fateev}.  (The spin is rescaled $S^z_{lat}(x)=0.783(3)S^z_{cont}(x)$ from the continuum to the lattice model~\cite{Destri,Delfino}.)  The relative mass of these bound states related by the $E_8$ symmetry at $h^x_c$, are presented in Fig.~\ref{2}D, again with good agreement in the analytically exact limit, see Tab.~\ref{tab:msw}~\cite{Zamolodchikov}.

\begin{table}[htbp] 
\begin{center}
  \begin{tabular}{c c c c c c c}
       $m_2/m_1$ & $m_3/m_1$ & $m_4/m_1$&$m_5/m_1$ &$m_6/m_1$ &$m_7/m_1$ & $m_8 / m_1$ \\ \hline
       1.618 &  1.989 &  2.405 &  2.956 &  3.218 &  3.891 &  4.783\\  \hline 
         \end{tabular}
  \caption{Analytically predicted mass ratios from Ref.~\onlinecite{Zamolodchikov}.   These numbers result from evaluation of simple trigonometric expressions (e.g., $m_2/m_1 = 2 \cos \pi/5$) that arise as eigenvalues of a matrix constructed from roots of the Lie algebra $E_8$.}   \label{tab:msw}
\end{center}
\end{table}

The deviation of the asymptotic expansion is slightly larger for higher bound states.  However, the deviation for large longitudinal fields ($h^z \lesssim h^x_c$) is fairly small, indicating influence of criticality up to very strong longitudinal fields.  Note also that the uncertainty of our results increase with decreasing longitudinal field strength when the bound state masses move closer, due to our fixed energy resolution.  For future reference we also present results as a function of the transverse field at longitudinal field $h^z=0.035$ meV present in $\textrm{CoNb}_2\textrm{O}_6$ around $h^x=h^x_c$, see Fig.~\ref{2}E.  Good agreement for the energies of the bound states are obtained with earlier numerical work using the \emph{Truncated Free Fermion Space Approach}, cf. Fig. 5 of Ref.~\onlinecite{Fonseca}.  Stronger longitudinal field will increase the minimum gap and move it to stronger transverse fields, but the gap increases slower away from its minimum value.  Also note that the minimum for higher bound state masses occurs for a lower transverse field.  The relative masses increase linearly around $h^x_c$, see Fig.~\ref{2}F, with a steeper slope for higher bound state masses and lower longitudinal fields.  
\begin{figure}[htbp]
  \begin{center}
    \subfigure{\includegraphics[width=42mm]{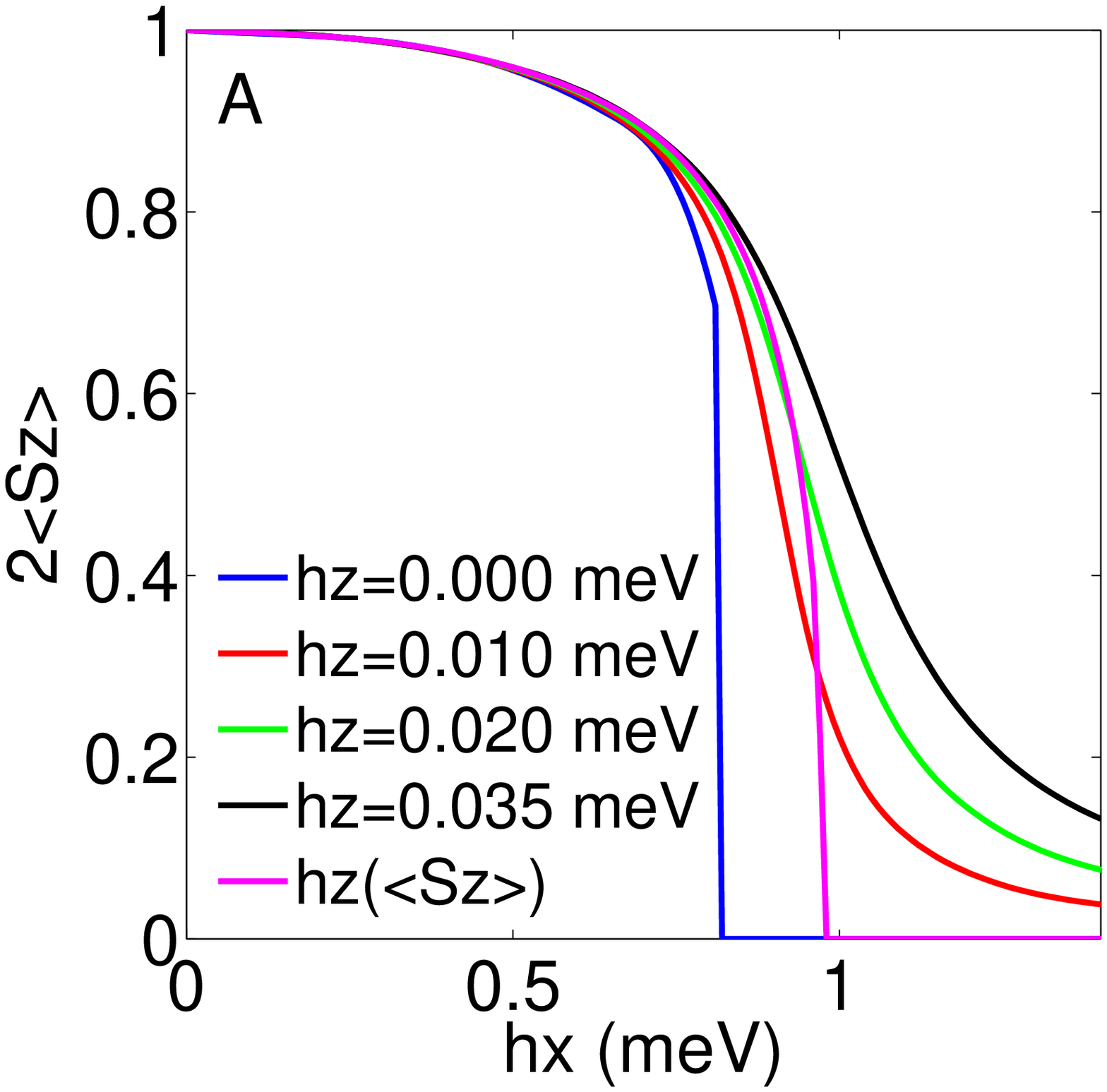}}
        \subfigure{\includegraphics[width=42mm]{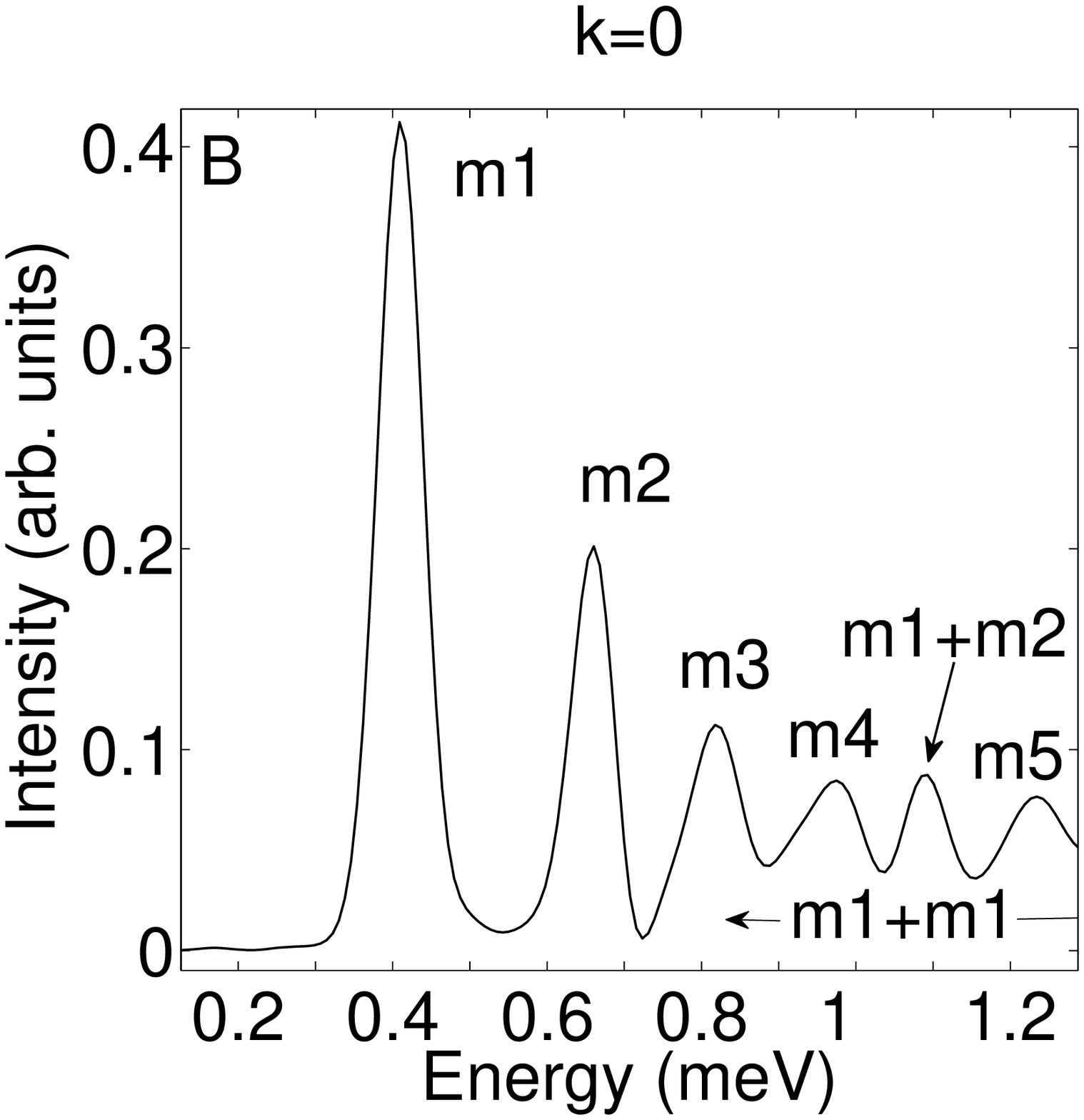}}\\
            \subfigure{\includegraphics[width=42mm]{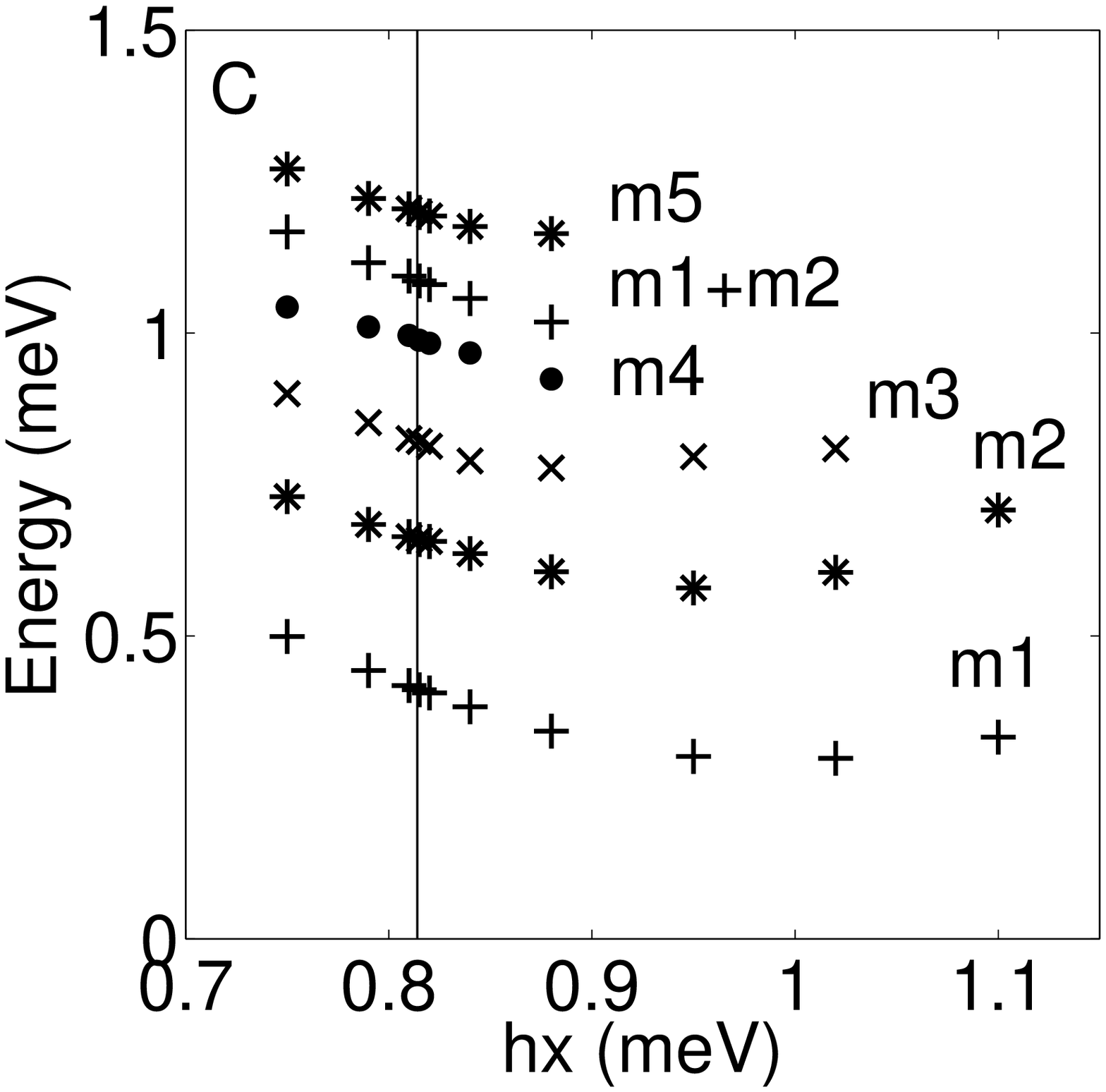}}
                \subfigure{\includegraphics[width=42mm]{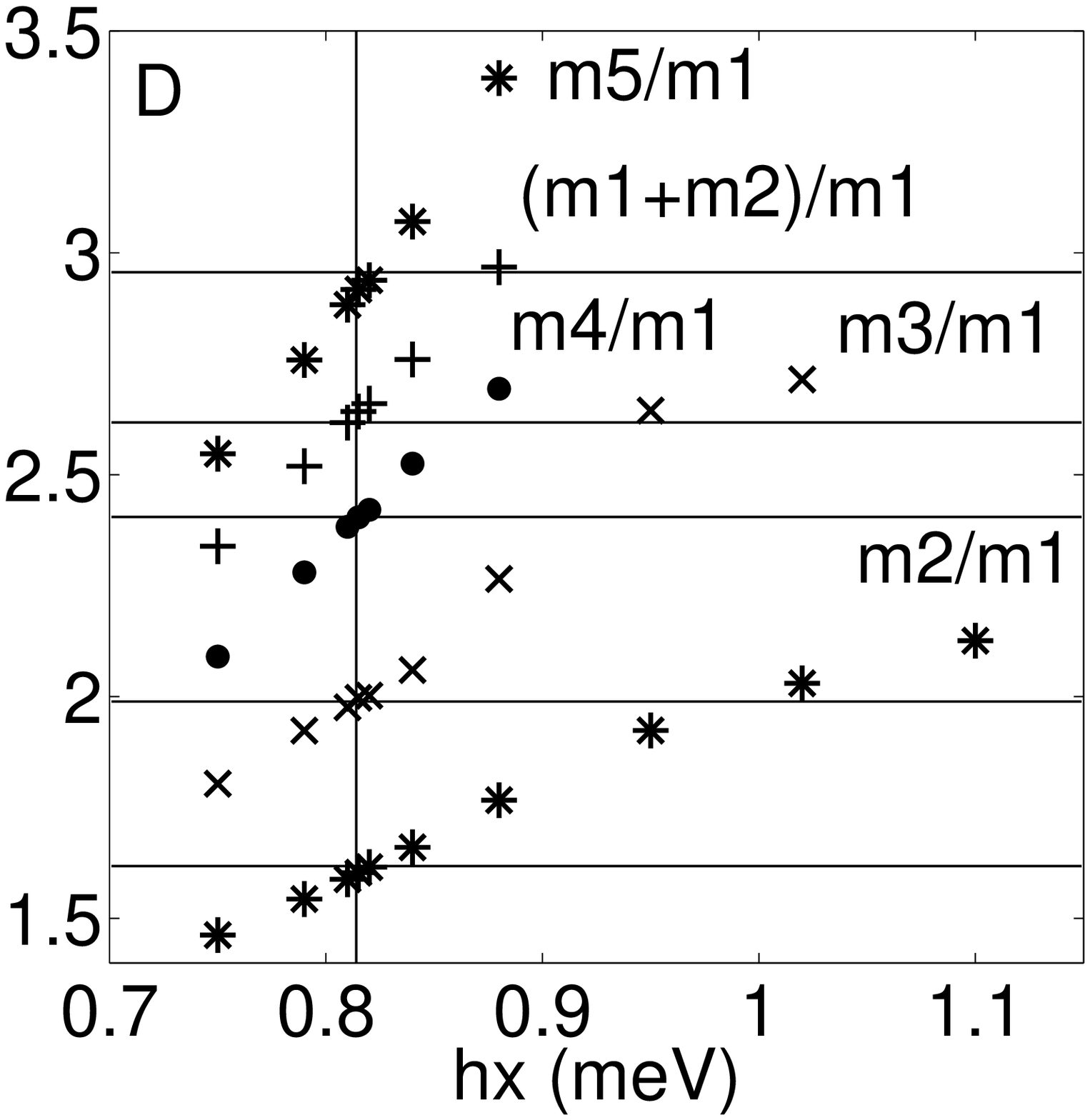}}
    \caption{(Color online) The full Hamiltonian describing $\textrm{CoNb}_2\textrm{O}_6$.  (A) Magnetization comparison between weakly coupled chains and chains in different constant longitudinal field.  (B) The cross section of the dynamical structure function for $h^x_c$ at zero momentum showing the masses of the first five bound states and two bound state pairs. (C) The bound state masses as a function of $h^x$. The minimum gap is above $h^x_c$ and the bound state mass minimum decreases with increasing mass. (D) The ratio of the bound state masses varies linearly around $h^x_c$ and goes through the analytically calculated values at $h^x_c$.}
    \label{4}
  \end{center}
\end{figure}

Finally we turn to the more accurate microscopic model of $\textrm{CoNb}_2\textrm{O}_6$ Eq.~\eqref{ph} with values of the coupling constants presented above.  The QCP at zero longitudinal field for this model is moved to a slightly weaker field $h_c^x\approx 0.814$ meV, see Fig.~\ref{4}A from ground state simulations with TEBD, due to the addition of the ferromagnetic XX-term.  The longitudinal field strength from weakly coupled chains $h^z(\langle S^z\rangle)$ is to a good approximation constant past $h_c^x$, see Fig.~\ref{4}A.  At vanishing magnetization this is not true, but here the 1D approximation of the 3D material is breaking down anyway.  The dynamical structure function at $h_c^x$, not presented here, shows a flattening of the kinetic bound state and a more prominent lowest bound state dispersion.  The cross section at zero momentum, see Fig.~\ref{4}B, has the same characteristics as the one for the quantum Ising chain, with fairly unaltered spectral weights.  The relative weight of the bound state continuum is still largest around $h_c^x$, making this region even more interesting for experiments.  A more careful analysis of the bound state masses, see Fig.~\ref{4}C, reveals a small rescaling of both axes to around 90\% of their previous values. This overall scaling does not affect the mass ratios, see Fig.~\ref{4}D.  Again they follow straight lines, see Fig.~\ref{4}D, and pass the analytical values at $h_c^x$, exactly as they do for the quantum Ising chain, rather than approaching it by bending as suggested by the extrapolation of the experimental data in Ref.~\onlinecite{Coldea}. Additional interactions irrelevant at low field and not treated here might explain the bending if it is confirmed by higher resolution data, but our results suggest higher resolution data will show that the mass ratios indeed go through the analytical values at the critical field if the model used here (and previously~\cite{Coldea,LeeBalents}) is a good one for $\textrm{CoNb}_2\textrm{O}_6$ past $h_c^x$. 

To conclude, we have investigated the effects of integrability near the Ising QCP and evaluated how far away the features extend and how robust they are to additional interactions.  We have shown that the bound state continuum should carry comparable spectral weight to the higher bound states.  The microscopic 1D model of $\textrm{CoNb}_2\textrm{O}_6$ treated here is able to reproduce the experimental data far away from criticality well.  When moved close to the QCP, the model still has the characteristics of the $E_8$ symmetry, with the mass ratios following straight lines through the analytical values, even better than the extrapolated experimental data suggests.  Future experiments with improved resolution on $\textrm{CoNb}_2\textrm{O}_6$ should detect higher bound state signatures to confirm the effects of integrability and the bound state continuum modeling confinement dynamics around the 1D QCP.

   
The authors thank R. Coldea for very useful correspondence.  This work was supported by a grant from the Army Research Office with funding from the DARPA OLE program and by the Knut and Alice Wallenberg foundation (J.~K.).

\end{document}